\documentclass[seceq]{ptptex}

\usepackage{graphicx}
\usepackage{wrapft}

\preprintnumber[80mm]{
\hfill\today}

\title{
Perturbative Approach to the Gravitational Lensing by a
Non-spherically Distorted 
Compact Object
}

\author{
Amano \textsc{Saijo}
and Masumi \textsc{Kasai}\footnote{E-mail: kasai@phys.hirosaki-u.ac.jp}
}
\inst{
Graduate School of Science and Technology, Hirosaki University,\\
Hirosaki 036-8561, Japan
}

\recdate{
}

\abst{
  We investigate the gravitational lens effect caused by a
  non-spherically distorted compact object. The non-spherical property
  of the gravitational potential is modeled by a quadrupole moment.
  Under the assumption that the quadrupole contribution is small, we
  solve perturbatively the lens equation and obtain the image
  positions and the amplification factors.  We show that the
  separation angle of two major images is only slightly changed by the
  existence of the quadrupole contribution, whereas the difference of
  the amplification factors may be significantly modified.  Our
  results indicate that even a tiny non-spherical distortion of the
  lens potential may cause significant amount of flux anomalies in the
  lensed images.
  }

\begin{document}
\maketitle

\section{Introduction}

The gravitational lens is an important tool in astrophysics for probing
mass distributions and determining the cosmological
parameters.\cite{esf}  
Although previous studies of gravitational lensing have been mostly based on
simple, spherically symmetric lens models,
some generalizations which include rotation\cite{rf:2}\tocite{ak}
 and higher-order general
relativistic effects\cite{eoak,aky} have also been developed. 
In this paper, we investigate another generalization, i.e., the effect
of non-spherical distortion of the gravitational potential in a
compact lens object on the gravitational 
lensing. 

A pioneering work on the non-spherically deformed compact lens was 
done by Asada.\cite{asada}  
He studied analytically the gravitational lensing caused by a
non-spherically deformed star, where the non-spherical property of the
gravitational potential was modeled by a quadrupole moment. 
He employed a rigorous analytic approach and presented solutions of
the image positions for a source on the principal axis, an expression of the caustics and the critical curves. 
However, because of the highly non-linear nature of the lens
equation, general solutions of the image positions for a off-axis
source and their amplification factors have not been presented. 

In this paper, we take a more practical approach, namely the
perturbative approximation, and solve analytically the lens equation
and obtain the approximate solutions up to  the lowest order of the
quadrupole moment. 
We also calculate the Jacobian of the lens mapping and indicate how
the quadrupole moment changes the amplification factors.

\section{Lens equation}

In this section, 
we summarize the lens equation for a compact object with a quadrupole
moment. See also Asada\cite{asada}  for detail. 
The gravitational potential of a non-spherically deformed compact
object is modeled by use of a monopole and a quadrupole moment: 
\begin{equation}
  \phi = \phi_0 + \phi_2 = -\frac{GM}{r} -
  \frac{G}{2}\left(3\frac{x^i x^j}{r^5} - \frac{\delta^{ij}}{r^3}
    \right) I_{ij}, 
\end{equation}
where
\begin{equation}
  M = \int \rho\, d^3x,\quad
  I_{ij} = \int \rho\, x_i x_j d^3 x, 
\end{equation}
and $\rho$ is the mass density of the lens object.
Using the above form of the potential, the 
deflection angle 2-vector $\boldsymbol\alpha$ is calculated as follows:
\begin{equation}
  \alpha^i = \frac{4GM}{c^2}\frac{\xi^i}{|{\boldsymbol\xi}|^2}
  + \frac{8G}{c^2} \left(
    2Q_{jk}\frac{\xi^j\xi^k \xi^i}{|{\boldsymbol\xi}|^6} -
     Q_{ij}\frac{\xi^j}{|{\boldsymbol\xi}|^4}\right), 
\end{equation}
where the 2-dimensional vector
$\boldsymbol \xi$ denotes the image position, and 
\begin{equation}
  Q_{ij} = \int\rho\,\left(X_iX_j -
    \frac{1}{2}\delta_{ij}|{\boldsymbol X}|^2
\right) d^3 X
\end{equation}
denotes the trace-free quadrupole moment.
Without loss of generality, we can assume the normalized quadrupole moment is
 diagonalized as follows,
\begin{equation}
  \tilde{Q}_{ij} \equiv \frac{c^2 D_S}{2GM^2 D_L D_{LS}} Q_{ij}
  = \left(
\begin{array}{cc}
e & 0\\
0 & -e\\
\end{array}
\right), 
\end{equation}
where $D_S, D_L, D_{LS}$ are the angular diameter distances from the
observer to the source, from the observer to the lens, and from the
lens to the source, respectively. 
Hereafter, we assume $e>0$. 
Finally, the lens equation is
\begin{equation}\label{eq:leq1}
  \beta_x = x - \frac{x}{x^2 + y^2} - e
  \frac{(x^2-3y^2) x}{(x^2+y^2)^3}, 
\end{equation}
\begin{equation}\label{eq:leq2}
  \beta_y = y - \frac{y}{x^2 + y^2} - e
  \frac{(3x^2-y^2) y}{(x^2+y^2)^3}, 
\end{equation}
where ${\boldsymbol\beta}=(\beta_x,\beta_y)$ and 
 ${\boldsymbol\theta}=(x, y)$ are the source and the image positions,
 respectively, normalized by the Einstein radius $\theta_E =
 \sqrt{4GMD_{LS}/c^2 D_L D_S}$. 

\section{Perturbative approach}

A comprehensive, algebraic study of the above lens equation
Eqs.~(\ref{eq:leq1}) and (\ref{eq:leq2}) was done by Asada\cite{asada} using
polar coordinates. He showed the lens equation for this type as a
single real 10th-order algebraic equation and gave analytic solutions
for a source located exactly on the principal axes.  
He also estimated the order of magnitude of the normalized quadrupole
moment as
\begin{equation}
  e  \sim  10^{-5} \left(\frac{M_{\odot}}{M}\right)
  \left(\frac{R}{10^6 \,\mbox{km}}\right)^3
  \left(\frac{10^7 \,\mbox{km}}{R_{\rm{E}}}\right)^2
  \left(\frac{v}{10\,{\mbox{km s$^{-1}$}}}\right)^2, 
\end{equation}
where $R$ denotes the typical size of a lens object, $R_{\rm{E}}$
is the Einstein ring radius, and $v$ is a rational surface velocity
on the equatorial plane of the lens star. 
Then, for a nearby solar-type star at 10 pc, i.e., $R\sim 10^6$ km,
$R_{\rm{E}} \sim 10^7$ km,  with a rotational
velocity of $v\sim 10$ km s$^{-1}$ which is much faster than that of
the Sun,  the estimated value is $e \sim 10^{-5}$ , which is
sufficiently smaller 
than unity. 
 
Because
of the higher polynomial nature of the lens equations, however, general
analytic solutions for a source not on the axes have not been given
yet (or cannot be given for more than 4 images). Also, the physical
properties of the lensed images, such as the image amplification
factors, are not yet investigated.

Instead of such an exact algebraic treatment, in this paper, we take another, a
more tractable approach to the quadrupole lens equations.  
Using the fact $0<e\ll 1$, 
we employ a
perturbative approach to the lens equations and obtain approximate
solutions to the image positions $(x,y)$, up to the lowest order of
the eigenvalue $e$ of the quadrupole moment. 

Concerning the perturbative approach to the gravitational lensing, 
for example, 
Alard\cite{a1}\tocite{a3}  wrote a series of papers in a slightly different
context from ours. 
His main interest is the non-spherical perturbations of extended lens
models and elongated arc images, whereas in this paper we concentrate our
attention on a non-spherically distorted compact lens model and multiple
images. 

We start from the following set of the lens equation, which is
obtained by clearing the fraction of Eqs.~(\ref{eq:leq1}) and
(\ref{eq:leq2}): 
\begin{equation}\label{lenseq1}
  (x^2+y^2)^2 \left\{
    (x^2+y^2)(\beta_x - x) + x\right\} = -e (x^2-3 y^2) x, 
\end{equation}
\begin{equation}
  \label{lenseq2}
   (x^2+y^2)^2 \left\{
    (x^2+y^2)(\beta_y - y) + y\right\} = -e (3 x^2-y^2) y. 
\end{equation}
Setting $e=0$, we obtain the zeroth-order solutions, 
\begin{equation}
  \label{0th-sols}
  x=x_0^{\pm} \equiv f^{\pm} \beta_x, \quad y=y_0^{\pm} \equiv f^{\pm} \beta_y, 
\end{equation}
where 
\begin{equation}
  f^{\pm} \equiv \frac{1\pm \sqrt{1+4\beta^{-2}}}{2}, 
\end{equation}
and $\beta=\sqrt{\beta_x^2+\beta_y^2}$. 
The above solutions are just the solutions for a point mass lens. 
We may also find a trivial solution $(x, y)=(0,0)$, but it is inadequate
because the denominators in the original lens equation
Eqs.~(\ref{eq:leq1}) and (\ref{eq:leq2}) vanish. 

Next, we put the following form into Eqs.~(\ref{lenseq1}) and
(\ref{lenseq2}), 
\begin{equation}
  x=x_0 + x_1, \quad y=y_0 + y_1, 
\end{equation}
then, up to the linear order of $x_1$ and $y_1$, we obtain 
\begin{equation}
  (2x_0x _1 + 2 y_0 y_1)(\beta_x-x_0)- (x_0^2+y_0^2) x_1 + x_1 =
  -e\frac{(x_0^2-3y_0^2)x_0}{(x_0^2+y_0^2)^2}, 
\end{equation}
\begin{equation}
 (2x_0x _1 + 2 y_0 y_1)(\beta_y-y_0)- (x_0^2+y_0^2) y_1 + y_1 =
   -e\frac{(3x_0^2-y_0^2)y_0}{(x_0^2+y_0^2)^2}. 
\end{equation}
The solutions are expressed in terms of $\beta_x$ and $\beta_y$ as
follows: 
\begin{equation}\label{majorx}
  x^{\pm}=x_0^{\pm}+x_1^{\pm} = f^{\pm}\beta_x +
  e\frac{(4\beta_x^2-3\beta^2)(f^{\pm})^2-1}
  {\beta^2(f^{\pm}\beta^2+1)(f^{\pm}\beta^2+2)}\beta_x,
\end{equation}
\begin{equation}\label{majory}
  y^{\pm}=y_0^{\pm}+y_1^{\pm} = f^{\pm}\beta_y +
  e\frac{(3\beta^2-4\beta_y^2)(f^{\pm})^2+1}
  {\beta^2(f^{\pm}\beta^2+1)(f^{\pm}\beta^2+2)}\beta_y.
\end{equation}

In the case of a point mass lens model ($e=0$), the number of the images is
two. As shown by Asada,\cite{asada}  the number of the images by a
quadrupole lens ($e\neq 0$), which
depends on the value of $e$ and the position of the source, is
generally more than two, in most cases four. 
We can obtain another, new solutions for $e>0$ case, which are not
existent for a point mass lens case,  as a perturbation
around the trivial solution $(x,y)=(0,0)$, namely,   
\begin{equation}
  x=0 + x_1, \quad y=0 + y_1. 
\end{equation}
Up to the lowest non-trivial order, 
the lens equation is  
\begin{equation}\label{eq311}
  (x_1^2+y_1^2)^2 x_1=-e(x_1^2-3y_1^2) x_1,
\end{equation}
\begin{equation}\label{eq312}
  (x_1^2+y_1^2)^2 y_1=-e(3x_1^2-y_1^2) y_1. 
\end{equation}
In the case of $x_1\neq 0, y_1\neq 0$, Eqs.~(\ref{eq311}) and
(\ref{eq312}) imply
$x_1^2 = -y_1^2$, which does not have a real root. 
In the case $x_1\neq 0, y_1=0$,  Eq.~(\ref{eq311})  implies
$x_1^2 = -e$, which again does not have a real root since we have
assumed $e>0$. Finally in the case $x_1=0, y_1\neq 0$,
Eq.~(\ref{eq312}) 
 reduces to $y_1^2=e$, from which we obtain
\begin{equation}
  x = x_1 = 0, \quad y=y_1 = \pm \sqrt{e}. 
\end{equation}
The above solutions are always on the $y$-axis, independent of the
source position $(\beta_x,\beta_y)$, and order of $O(\sqrt{e})$, not
$O(e)$. Therefore, we perform one more iteration and obtain slightly
higher order solutions in the following form, 
\begin{equation}
  x=0+x_2, \quad y=0+y_1 + y_2 = \pm \sqrt{e} + y_2. 
\end{equation}
Up to the lowest order of $x_2$ and $y_2$, the lens equation is now
\begin{equation}
  y_1^6\beta_x + y_1^4 x_2 = 3ey_1^2 x_2, 
\end{equation}
\begin{equation}
  y_1^6 (\beta_y -y_1) +  5y_1^4 y_2 = 3 e y_1^2 y_2, 
\end{equation}
and the solution is
\begin{equation}\label{minor}
  x = x_2 = \frac{1}{2} e \beta_x, \quad
 y = y_1 + y_2 = \pm \sqrt{e}\left(1+\frac{1}{2}e\right)
  - \frac{1}{2} e \beta_y. 
\end{equation}
These images always appear very close to the $y$-axis, are very dim
because the amplitude of the amplification factor is $O(e^2)$, and
disappear when $e\rightarrow 0$.  We call them as the ``minor''
images, and may safely neglect them in the analysis of the image
amplifications.

\begin{table}[htb]
\caption{Comparison of our approximate solutions with numerical ones
  for $\beta_x=0$}
\label{table1}
\centering
 \begin{tabular}{cccrrrrrr}
 \hline\hline
 $e$ & $\beta_x$ & $\beta_y$ &
 $x_{\rm num}$ & $x_{\rm appr}$ &
 (1) $y_{\rm num}$ & (2) $y_{\rm appr}$ &
 $|(2)-(1)|$ \\
  \hline
  0.01 & 0.0 & 0.2 & 0.0 & 0.0  & 1.10087 & 1.10091 & $3.9\times 10^{-5}$ \\
  &     &     & 0.0 & 0.0  & -0.89881 & -0.89891 & $1.0\times 19^{-4}$ \\
  &     &     & 0.0 & 0.0  & 0.09950 & 0.099500 & $3.7\times 10^{-6}$ \\
  &     &     & 0.0 & 0.0  & -0.10157 & -0.10150 & $6.8\times 10^{-5}$ \\
  \hline
  0.01 & 0.0 & 0.5 & 0.0 & 0.0  & 1.27780 & 1.27782 & $1.8\times 10^{-5}$ \\
  &     &     & 0.0 & 0.0  & -0.77262 & -0.77282 & $2.0\times 10^{-4}$ \\
  &     &     & 0.0 & 0.0  & 0.09809 & 0.09800 & $8.5\times 10^{-5}$ \\
  &     &     & 0.0 & 0.0  & -0.10327 & -0.10300 & $2.7\times 10^{-4}$ \\
  \hline
  0.02 & 0.0 & 0.2 & 0.0 & 0.0  & 1.09668 & 1.09684 & $1.6\times 10^{-4}$ \\
  &     &     & 0.0 & 0.0  & -0.89241 & -0.89284 & $4.2\times 10^{-4}$ \\
  &     &     & 0.0 & 0.0  & 0.14084 & 0.14084 & $2.5\times 10^{-7}$ \\
  &     &     & 0.0 & 0.0  & -0.14510 & -0.14484 & $2.7\times 10^{-4}$ \\
  \hline
  0.02 & 0.0 & 0.5 & 0.0 & 0.0  & 1.27479 & 1.27486 & $7.3\times 10^{-5}$ \\
  &     &     & 0.0 & 0.0  & -0.76402 & -0.76486 & $8.4\times 10^{-4}$ \\
  &     &     & 0.0 & 0.0  & 0.13802 & 0.13784 & $1.8\times 10^{-4}$ \\
  &     &     & 0.0 & 0.0  & -0.14878 & -0.14784 & $9.5\times 10^{-4}$ \\
  \hline
 \end{tabular}
\end{table}

In order to check the accuracy of our approximate solutions
Eqs.~(\ref{majorx}), ({\ref{majory}), and (\ref{minor}), 
we compare them with numerical solutions. Table \ref{table1} and
\ref{table2} show the result for some sets of typical values of the
parameters. 
Table \ref{table1} shows the calculated values for the source on the
$y$-axis, i.e., $\beta_x=0$, and Table \ref{table2} for the source on
the $x$-axis, $\beta_y=0$.  The maximal error is about $10^{-4}$, which is
the same order of magnitude as $O(e^2)$. 

\begin{table}[htb]
\centering
\caption{Comparison of our approximate solutions with numerical ones
  for $\beta_y=0$}
\label{table2}
 \begin{tabular}{cccrrrrrr}
 \hline\hline
 $e$ & $\beta_x$ & $\beta_y$ &
 (1) $x_{\rm num}$ & (2) $x_{\rm appr}$ &
 $|(2)-(1)|$ &
 (3) $y_{\rm num}$ & (4) $y_{\rm appr}$ &
 $|(4)-(3)|$ \\
  \hline
  0.01 & 0.2 & 0.0 &
  1.10902 & 1.10906 & $3.8\times 10^{-5}$ &
  0.0 & 0.0 & 0.0 \\
  &     &     &
  -0.91097 & -0.91106 & $9.7\times 10^{-5}$ &
  0.0 & 0.0 & 0.0 \\
  &     &     &
  0.00102 & 0.00100 & $2.0\times 10^{-5}$ &
  0.10048 & 0.10050 & $1.7\times 10^{-5}$ \\
  &     &     &
  0.00102 & 0.00100 & $2.0\times 10^{-5}$ &
  -0.10048 & -0.10050 & $1.7\times 10^{-5}$ \\
  \hline
  0.01 & 0.5 & 0.0 &
  1.28372 & 1.28373 & $1.8\times 10^{-5}$ &
  0.0 & 0.0 &  0.0 \\
  &     &     &
  -0.78855 & -0.78873 & $1.9\times 10^{-4}$&
  0.0 & 0.0 & 0.0 \\
  &     &     &
  0.00254 & 0.00250 & $3.8\times 10^{-5}$ &
  0.10035 & 0.10050 & $1.5\times 10^{-4}$ \\
  &     &     &
  0.00254 & 0.00250 & $3.8\times 10^{-5}$ &
  -0.10035 & -0.10050 & $1.5\times 10^{-4}$ \\
  \hline
  0.02 & 0.2 & 0.0 &
  1.11299 & 1.11314 & $1.5\times 10^{-4}$ &
  0.0 & 0.0 & 0.0 \\
  &     &     &
  -0.91676 & -0.91714 & $3.8\times 10^{-4}$ &
  0.0 & 0.0 & 0.0 \\
  &   &     &
  0.00208 & 0.00200 & $8.1\times 10^{-5}$&
  0.14281 & 0.14284 & $2.5\times 10^{-5}$ \\
  &     &     &
  0.00208 & 0.00200 & $8.1\times 10^{-5}$&
  -0.14281 & -0.14284 & $2.5\times 10^{-5}$ \\
  \hline
  0.02 & 0.5 & 0.0 &
  1.28662 & 1.28669 & $7.0\times 10^{-5}$&
  0.0 & 0.0 & 0.0 \\
  &     &     &
  -0.79598 & -0.79669 & $7.1\times 10^{-4}$ &
  0.0 & 0.0 & 0.0 \\
  &     &     &
  0.00516 & 0.00500 & $1.6\times 10^{-4}$ &
  0.14241 & 0.14284 & $4.2\times 10^{-4}$ \\
  &     &     &
  0.00516 & 0.00500 & $1.6\times 10^{-4}$ &
  -0.14241 & -0.14284 & $4.2\times 10^{-4}$ \\
  \hline
 \end{tabular}
\end{table}

\section{Amplification factor}

Once we express the solutions for the image positions $(x,y)$ in terms
of the source position $(\beta_x,\beta_y)$, we can directly calculate
the amplification factor from the Jacobian 
\begin{equation}
  A^{\pm} = \left| \det\frac{\partial(x^{\pm},y^{\pm})}
    {\partial(\beta_x,\beta_y)}\right|,  
\end{equation}
where $A^{+}$ and $A^{-}$ represent the amplification factors for the
images with positive and negative parity, respectively. 
The calculation is straightforward, and from Eqs.~(\ref{majorx}) and
(\ref{majory}) we obtain the following results up to the linear order
of $e$: 
\begin{equation}
  A^{+}=
  A_0^{+} 
  \left( 1 - e \frac{4(\beta_x^2-\beta_y^2)
      \left(\beta(\beta^2+6) -(\beta^2+4)\sqrt{\beta^2+4}\right)}
    {\beta^3(\beta^2+4)\left(\sqrt{\beta^2+4}+\beta\right)^2}
  \right), 
\end{equation}
\begin{equation}
  A^{-}=
  A_0^{-}
  \left( 1 - e \frac{4(\beta_x^2-\beta_y^2)
      \left(\beta(\beta^2+6)+(\beta^2+4)\sqrt{\beta^2+4}\right)}
    {\beta^3(\beta^2+4)\left(\sqrt{\beta^2+4}-\beta\right)^2}
  \right), 
\end{equation}
where 
\begin{equation}
    A_0^{\pm} \equiv 
    \frac{\left(\sqrt{\beta^2+4}\pm \beta\right)^2}{4\beta\sqrt{\beta^2+4}}   
\end{equation}
represent the amplification factors for a point mass lens, namely
$e=0$ case. 

We can also define the Pad\'e approximants for $A^{\pm}$ of order
$[0/1]$\footnote{
Generally speaking, the Pad\'e approximant of order $[m/n]$ is an
approximation of a function by a ratio of two polynomials of order $m$
and $n$. Then, for a function $A(x)$ whose Taylor expansion is given
by $A(x) \simeq A_0(1+ a_1 x+\cdots)$, the Pad\'e approximant of order
$[0/1]$ is simply $A_0/(1-a_1 x)$.  At the same order, Pad\'e
approximations are usually 
superior to Taylor expansions especially when functions contain
poles. }
 with respect to $e$ as follows: 
\begin{equation}\label{eq:ppade}
  A_{\rm P}^{+}\equiv
  A_0^{+} 
  \left( 1 + e \frac{4(\beta_x^2-\beta_y^2)
      \left(\beta(\beta^2+6)-(\beta^2+4)\sqrt{\beta^2+4}\right)}
    {\beta^3(\beta^2+4)\left(\sqrt{\beta^2+4}+\beta\right)^2}
  \right)^{-1}, 
\end{equation}
\begin{equation}\label{eq:mpade}
  A_{\rm P}^{-}\equiv
  A_0^{-}
  \left( 1 + e \frac{4(\beta_x^2-\beta_y^2)
      \left(\beta(\beta^2+6)+(\beta^2+4)\sqrt{\beta^2+4}\right)}
    {\beta^3(\beta^2+4)\left(\sqrt{\beta^2+4}-\beta\right)^2}
  \right)^{-1} .  
\end{equation}

\begin{table}[b]
\centering
\caption{Comparison of our approximate amplification factors with
    numerical ones}
\label{table3}
\begin{tabular}{ccccrrcrc}
  \hline\hline
  $e$ & $\beta_x$ & $\beta_y$ & parity & (1) $A_{\rm num}$ &
  (2) $A$ &$|(2)-(1)|/(1)$ & (3) $A_{\rm P}$ & $|(3)-(1)|/(1)$\\
  \hline
  0.01    & 0.0    & 0.2     & $+$ & 2.83741 &
  2.82454 & 0.45\% & 2.83847 & 0.04\% \\
          &        &         & $-$ & 2.37779 &
  2.32454 & 2.24\% & 2.37166 & 0.26\% \\
  \hline
  0.01    & 0.0    & 0.5     & $+$ & 1.56603 & 
  1.56568 & 0.02\% & 1.56609 & 0.00\% \\
          &        &         & $-$ & 0.65303 &
  0.64568 & 1.13\% & 0.65116 & 0.29\% \\
  \hline   
  0.02    & 0.0    & 0.2     & $+$ & 2.66004 &
  2.61173 & 1.82\% & 2.66404 & 0.15\% \\
          &        &         & $-$ & 2.87313 &
  2.61173 & 9.10\% & 2.83724 & 1.25\% \\
  \hline
  0.02    & 0.0    & 0.5     & $+$ & 1.54134 &
  1.53994 & 0.09\% & 1.54156 & 0.01\% \\
          &        &         & $-$ & 0.73395 &
  0.69994 & 4.63\% & 0.72434 & 1.31\% \\
  \hline
  0.01    & 0.2    & 0.0     & $+$ & 3.26482 &
  3.25015 & 0.45\% & 3.26619 & 0.04\% \\
          &        &         & $-$ & 1.78910 &
  1.75015 & 2.18\% & 1.78564 & 0.19\% \\
  \hline
  0.01    & 0.5    & 0.0     & $+$ & 1.61749 &
  1.61714 & 0.02\% & 1.61758 & 0.00\% \\
          &        &         & $-$ & 0.54292 &
  0.53714 & 1.06\% & 0.54170 & 0.22\% \\
  \hline
  0.02    & 0.2    & 0.0     & $+$ & 3.52534 &
  3.46296 & 1.77\% & 3.53232 & 0.20\% \\
          &        &         & $-$ & 1.59981 &
  1.46296 & 8.55\% & 1.58928 & 0.66\% \\
  \hline
  0.02    & 0.5    & 0.0     & $+$ & 1.64434 &
  1.64288 & 0.09\% & 1.64460 & 0.02\% \\
          &        &         & $-$ & 0.50371 &
  0.48288 & 4.14\% & 0.49971 & 0.80\% \\
  \hline
\end{tabular}
\end{table}
As shown in Table \ref{table3}, the Pad\'e approximants
Eqs.~(\ref{eq:ppade}) and (\ref{eq:mpade}) generally give the
approximate amplifications with better accuracy of about $1\%$ or
less. 
Hereafter, we use Eqs.~(\ref{eq:ppade}) and (\ref{eq:mpade}) as our
approximate formulae for the amplification factors. 

\section{Changes in
 the image properties}

For a point mass lens model, we have the following ``universal''
relations for the image properties. First, the sum of the two
image positions is
\begin{equation}\label{univ1}
  x^{+}_0 + x^{-}_0 = \beta_x, \quad y^{+}_0 + y^{-}_0 = \beta_y. 
\end{equation}
Namely, the sum of the image positions is always equal to the original
source position. 
Second, the difference of the amplification factors is always unity:
\begin{equation}\label{univ2}
 A_0^{\rm diff} \equiv A^{+}_0 - A^{-}_0 = 1. 
\end{equation}
This also means, for a point mass lens, the image with positive parity
is always brighter than that with negative parity. 
Just for reference, we also show that the image separation and the total
amplification are expressed in the following way: 
\begin{equation}
  \Delta x_0\equiv x^{+}_0 - x^{-}_0 = \frac{\sqrt{\beta^2+4}}{\beta} \beta_x, \quad
  \Delta y_0\equiv y^{+}_0 - y^{-}_0 = \frac{\sqrt{\beta^2+4}}{\beta} \beta_y,
\end{equation}
\begin{equation}
  A_0^{\rm tot}\equiv A^{+}_0 + A^{-}_0
  = \frac{\beta^2+2}{\beta\sqrt{\beta^2+4}}. 
\end{equation}

In this section, we investigate how the quadrupole moment changes the
above image properties. 
From Eqs.~(\ref{majorx}) and (\ref{majory}), we obtain the sum as 
\begin{equation}
  x^{+} + x^{-} = \beta_x - e \left(1+\frac{4\beta_y^2}{\beta^4}\right)
  \beta_x,\quad
  y^{+} + y^{-} = \beta_y + e \left(1+\frac{4\beta_x^2}{\beta^4}\right) \beta_y,  
\end{equation}
The change due to $e$ is getting larger as $\propto
4e\beta^{-2}$ for $|\beta| \ll 1$. 
Therefore, even if $|e|\ll 1$, the ``universal'' relation
Eq.~(\ref{univ1}) may be significantly broken if $|\beta| \ll 1$. 

On the contrary, the image separation is
\begin{equation}
  \Delta x \equiv x^{+} - x^{-} =
  \Delta x_0 \left\{
    1 - \left(
      \frac{2(\beta_x^2-\beta_y^2)}{\beta^2(\beta^2+4)}-1
        \right) e \right\},  
\end{equation}
\begin{equation}
  \Delta y \equiv y^{+} - y^{-} =
  \Delta y_0 \left\{
    1 - \left(
      \frac{2(\beta_x^2-\beta_y^2)}{\beta^2(\beta^2+4)}+1
        \right) e \right\}. 
\end{equation}
It is evident that, even if $|\beta| \ll 1$, the change due to $e$ in
the image separation is $\sim e\times O(1)$. Therefore, we may
conclude that the change due to $e$ in the image separation is small for $|e|\ll
1$.

Similar situation occurs in the analysis of the image amplification. 
From Eqs.~(\ref{eq:ppade}) and (\ref{eq:mpade}), the amplification
difference for $|\beta| \ll 1$ is
\begin{equation}\label{eq:diff}
  A^{\rm diff} \equiv A^{+}_P - A^{-}_P \simeq \frac{1+\frac{2}{\beta^2}e'}
  {1+ \frac{7}{2}e' - \left(\frac{2}{\beta}e'\right)^2}, 
\end{equation}
where 
\begin{equation}
  e' = \left( \left(\frac{\beta_x}{\beta}\right)^2
              -\left(\frac{\beta_y}{\beta}\right)^2
       \right) e. 
\end{equation}
The change due to $e$ is getting larger as $\propto
2e\beta^{-2}$ for $|\beta| \ll 1$. 
Therefore, even if $|e|\ll 1$, the ``universal'' relation
Eq.~(\ref{univ2}) may be significantly broken if $|\beta| \ll 1$.
As is already shown numerically in Table \ref{table3}, there is even a case
of $A^{\rm diff} < 0$ for $e=0.02, \beta_x=0.0, \beta_y=0.2$. 

The total amplification is
\begin{equation}\label{eq:5-10}
  A^{\rm tot}\equiv A^{+}_P + A^{-}_P \simeq \frac{A^{\rm tot}_0}
  {1+\frac{3}{2}e' - \left(\frac{2}{\beta}e'\right)^2}. 
\end{equation}
Compared to $A^{\rm diff}$, the change due to $e$ is relatively mild
for $|\beta| \ll 1$. 

We give some comments on the range of the parameters $e$ and
$\beta$. Since we have started from the linear perturbation with
respect to $e$, the range 
of validity is $0\leq e \ll 1$. If we assume additionally that the
maximal error in the amplification factors should be less than, say, $1\%$,
the range of validity may be $0 \leq e \lesssim 0.01$, which we can guess
from Table \ref{table3}. 
The ranges of $\beta_x$ and $\beta_y$ are more important. 
From Eqs.~(\ref{eq:diff}) and (\ref{eq:5-10}), we can obtain the constraint 
$|e'| \lesssim \beta^2$ and $|e'| \ll \beta$. 
If not, then the term of $O(e^2)$ in the denominators of
Eqs.~(\ref{eq:diff}) and (\ref{eq:5-10}) can make a dominant
contribution over the linear term, which is outside of the validity of
linear perturbation theory. 

\section{Summary}

In this paper, we have investigated the gravitational lens effect
caused by a non-spherical, compact object. 
The non-spherical property of the gravitational potential is expressed
by the quadrupole moment. 
Eqs.~(\ref{eq:leq1}) and (\ref{eq:leq2}) are the lens equations for a
compact lens object with a quadrupole moment. In these lens equations,
the effect of the non-spherical contribution is expressed by $e$, the
eigenvalue of the normalized quadrupole moment tensor $\tilde Q_{ij}$. 
 
Under the assumption that the quadrupole moment $e$ is small, we have
perturbatively solved the lens equation and obtained the solutions for
the image positions.  Eqs.~(\ref{majorx}) and (\ref{majory}) represent
the perturbative solutions for the positions of the ``major'' images,
which are reduced to the solutions for a point mass lens model in the
limit $e\rightarrow 0$.  We have also found two new perturbative
solutions for the ``minor'' images, which are given in
Eq.~(\ref{minor}).  The minor images are very dim, always appear very
close to the $y$-axis, near $(0,\pm \sqrt{e})$, and vanish in the
limit $e\rightarrow 0$. The accuracy of our approximate solutions is
numerically checked and summarized in Tables \ref{table1} and
\ref{table2} for some sets of parameters.  It is shown that the
maximal error in the image position is about $10^{-4}$, which is the
same order of magnitude as $O(e^2)$.

Then, we have calculated the amplification factors for the major
images from the Jacobian and obtained the Pad\'e approximants of them
in Eqs.~(\ref{eq:ppade}) and (\ref{eq:mpade}).  The accuracy of our
approximate formula for the amplification factor is numerically
estimated and summarized in Table \ref{table3} for some sets of
parameters.  It is shown that the typical relative error is $1\%$ or
less. 

For a point mass lens model, we know the simple ``universal''
relations for the image properties, Eqs.~(\ref{univ1}) and
(\ref{univ2}).  We have investigated how the quadrupole contribution
$e$ 
changes such ``universal'' relations. 
We have found that the change due to a non-zero $e$  in the image
separation is $\sim e\times O(1)$. 
However, the difference of the amplifications $A^{\rm diff} =
A^{+}-A^{-}$, which is always unity in the case of a point mass lens,
may be significantly changed due to the quadrupole contribution, which
is shown in Eq.~(\ref{eq:diff}). 

Mao and Schneider\cite{ms} discussed the ``anomalous'' flux ratio between the
images caused by a lens galaxy. 
The lens system they discussed was QSO B 1422 + 231. Whereas the 
well-known 
``simple'' lens models, such as a singular isothermal ellipsoid, could 
fit the observed image positions very accurately, they all failed to
obtain the observed flux ratios. The possibility they argued is that the
discrepancy between the observed and model-predicted flux ratios is due
to substructure or perturbation in the simple lens models, which only
slightly 
changes the image positions but significantly affects the flux
ratios. 
Since the lens model  in this paper is different from
that of Mao and Schneider,\cite{ms} 
we have no intention of directly applying our results to their lens
systems. Still, it should be noted that 
the results in our paper show some formal similarity.   
It indicates that even a tiny non-spherical distortion of the lens
potential may also cause significant amount of  flux
anomalies
 in the
lensed images, whereas it only slightly changes the image positions.

\end{document}